1990+1991

31 GHz

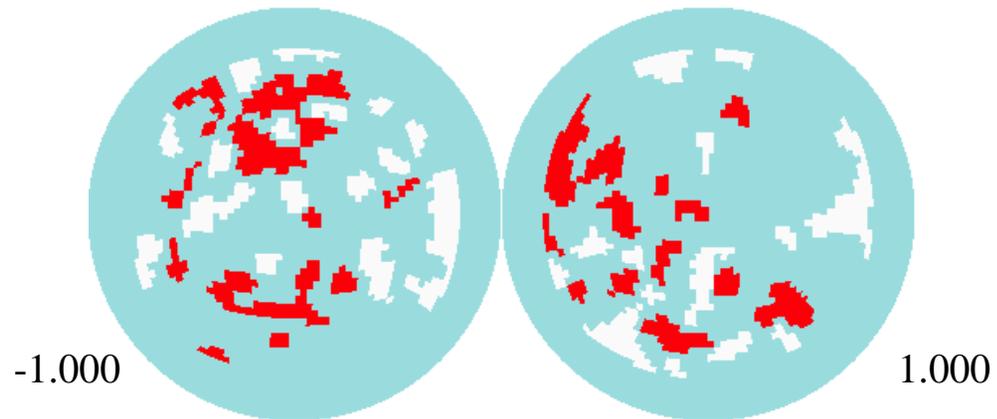

53 GHz

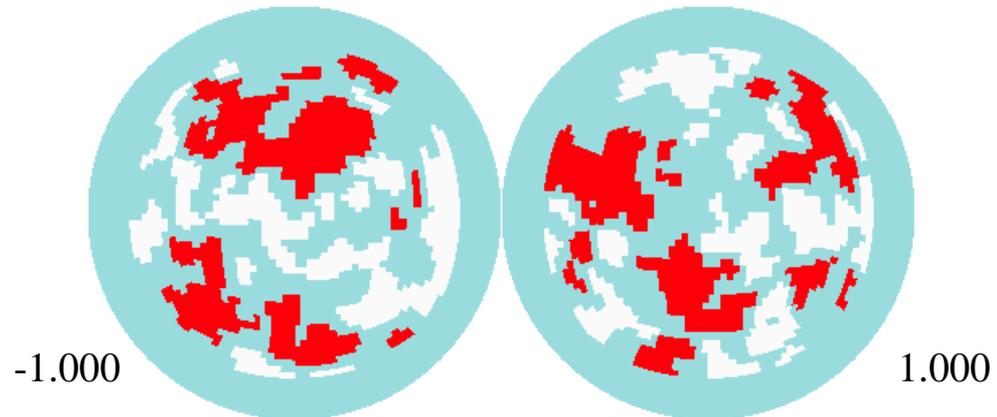

90 GHz

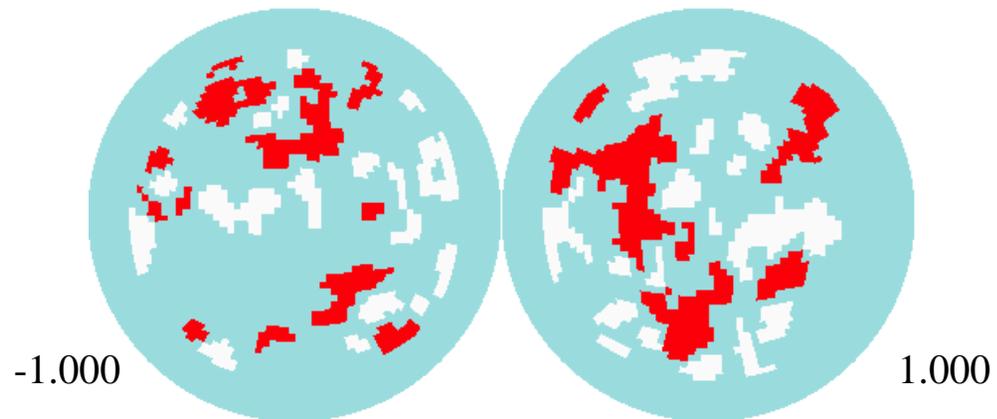

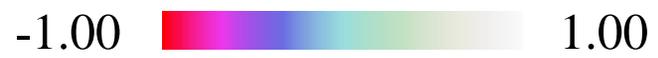

1990+1991

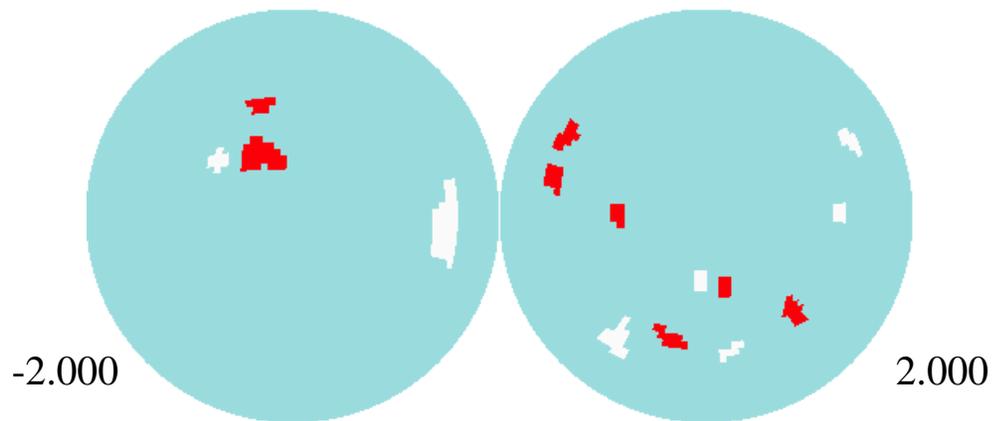

31 GHz

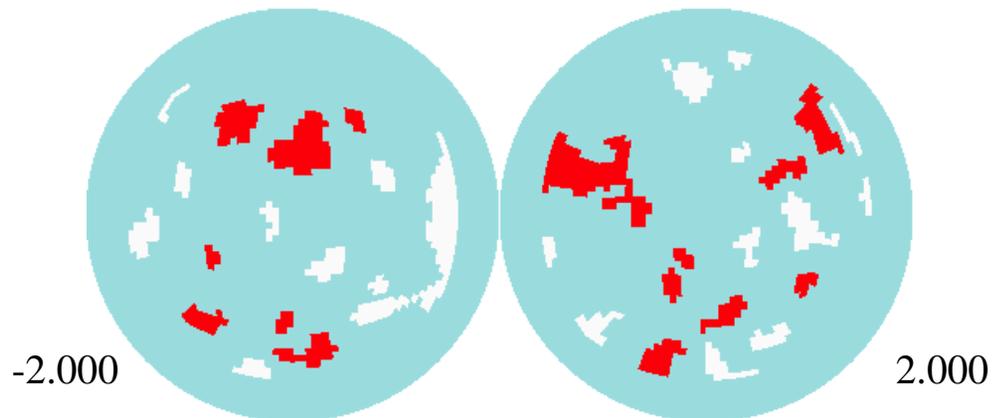

53 GHz

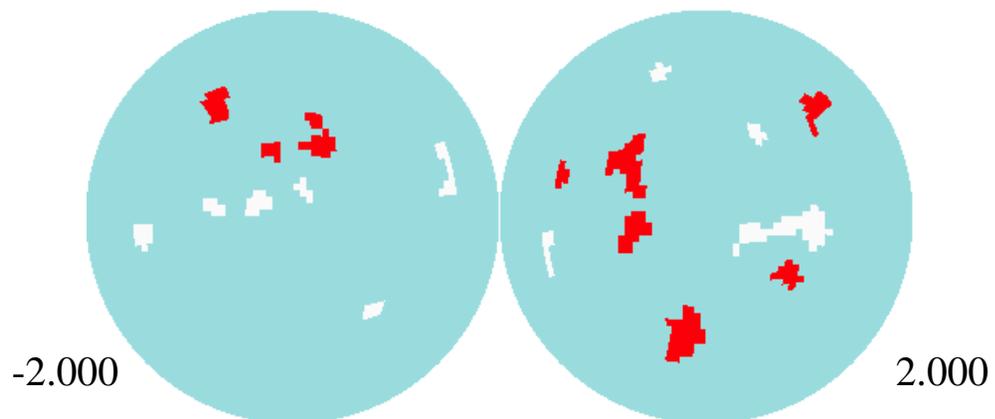

90 GHz

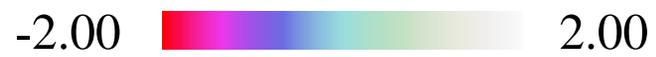

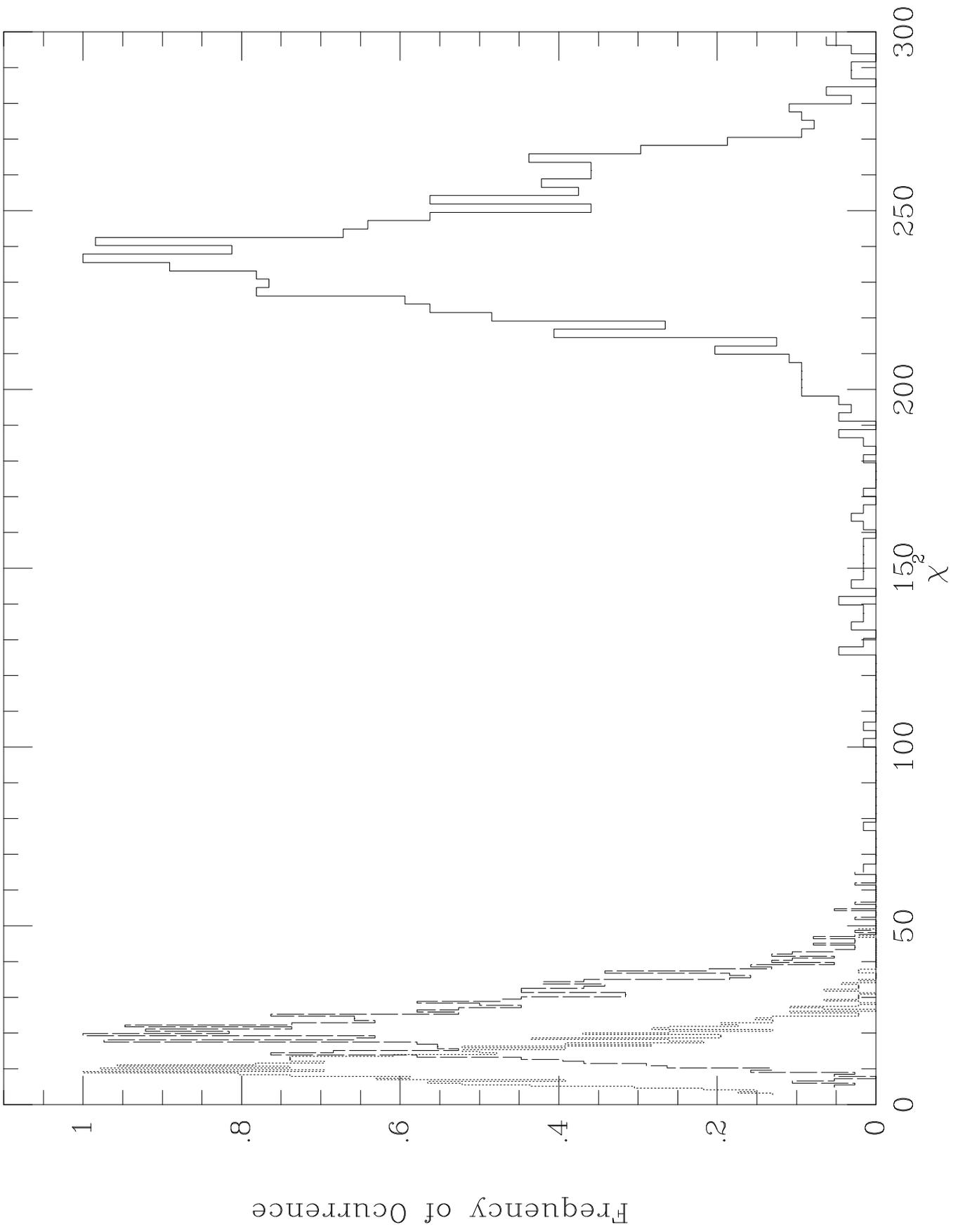

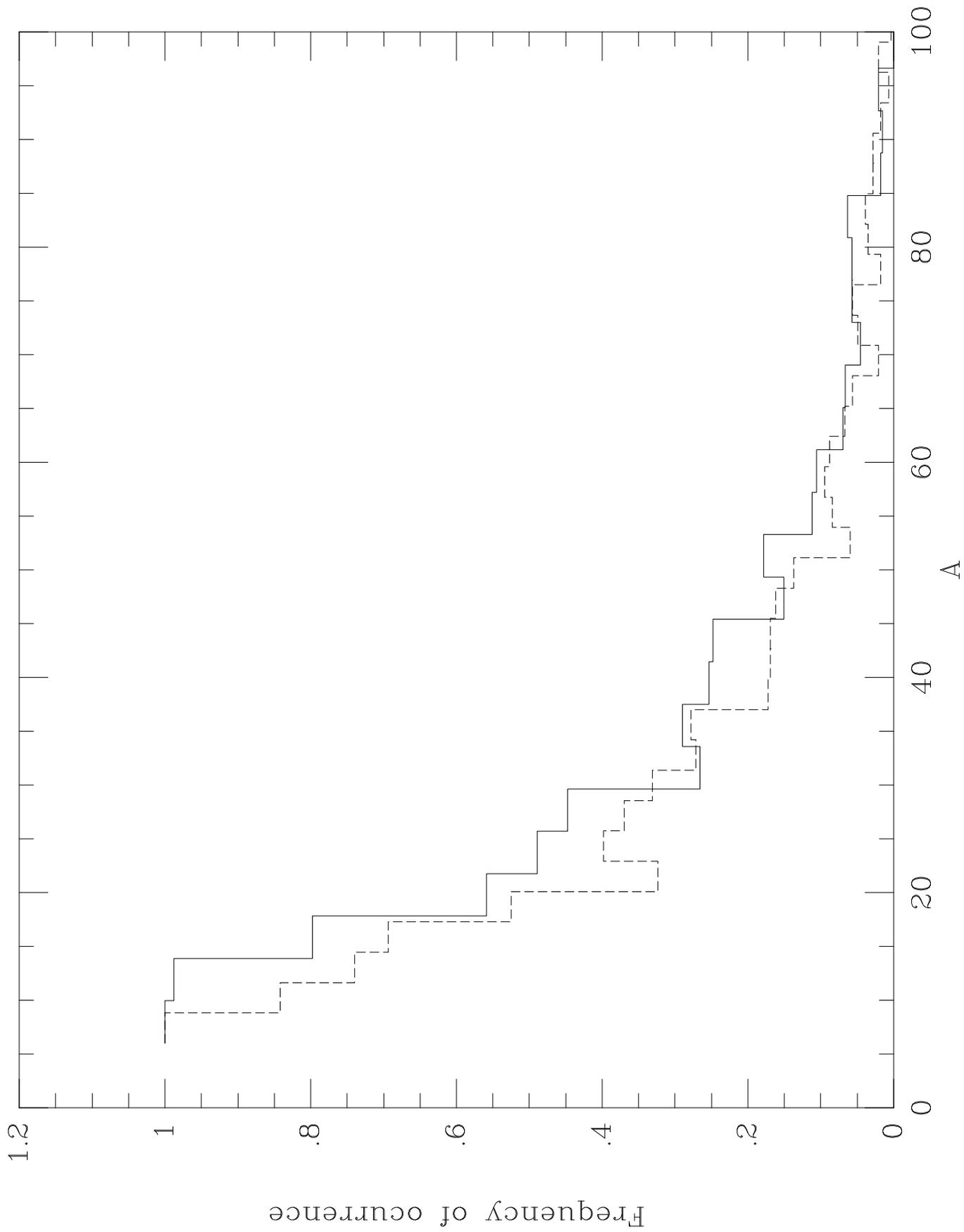

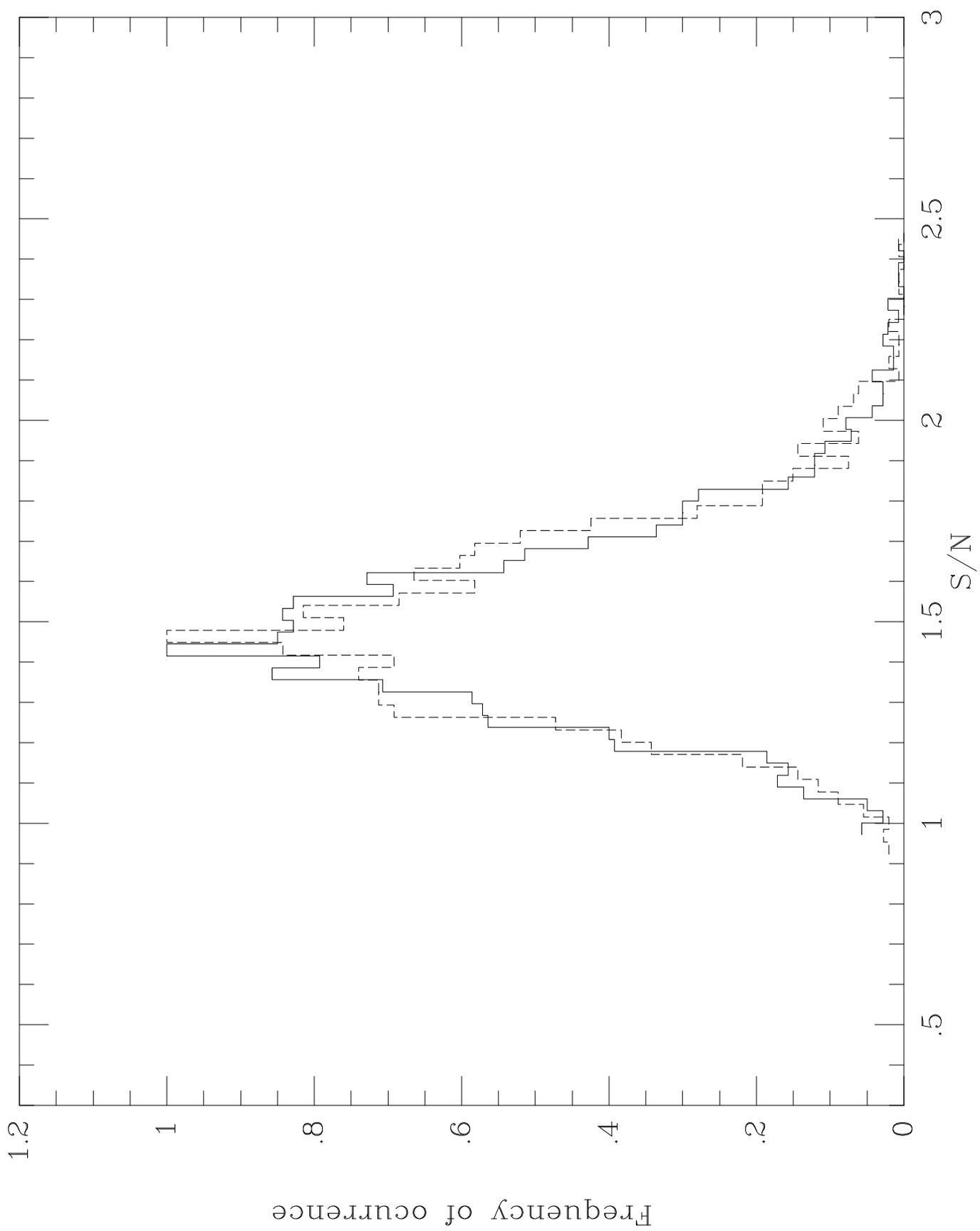

# Hot and Cold Spots
# in the First plus Second Year COBE/DMR Maps

L. Cayón and G. Smoot

Lawrence Berkeley Laboratory

Center for Particle Astrophysics

University of California, Berkeley, CA 94720

# ABSTRACT

Density perturbations at the decoupling epoch produce angular fluctuations in the temperature of the Cosmic Microwave Background (CMB) radiation that may appear as hot and cold spots. Observational data of the CMB includes instrumental noise in addition to the cosmological signal. One would like to determine which of the observed spots are produced by the noise and which correspond to signal. In this work we first present a statistical analysis of the first plus second year COBE/DMR map at 53 GHz that reveals the presence of cosmological signal in the data. The analysis is based on Harrison-Zeldovich Monte Carlo realizations and utilizes a generalized $\chi^2$ statistic. The method is applied to the number of spots and the fraction of the total area that appear above/below a certain value of the dispersion of the noise, including and excluding the quadrupole, giving $Q_{rms-PS} = 15^{+3}_{-6}, 18^{+5}_{-7} \mu K$ and $Q_{rms-PS} = 18^{+3}_{-4}, 21 \pm 6 \mu K$, at the 95% confidence level, respectively. The data taken by the COBE/DMR experiment during the first two years at three different frequencies (31, 53 and 90 GHz) are used to determine which of the spots observed at 53 GHz appear simultaneously in the other two channels. The significance of those spots is determined by comparison of their area and signal-to-noise with noise Monte Carlo simulations. We point out two cold spots and one hot spot at positions $(l, b) \approx (-99°, 57°), (-21°, -45°), (-81°, -33°)$ respectively, at the 95% confidence level.

*Subject headings:* cosmic microwave background: data analysis

1. Introduction

Temperature fluctuations of the CMB radiation at large angular scales are directly related to the matter density fluctuations at the last scattering surface (Sachs and Wolfe 1967). Direct observation of those fluctuations gives information on the seeds of the structures observed in the present Universe. Statistical analysis of the temperature of the CMB observed by the COBE/DMR experiment during 1990 revealed for the first time the presence of anisotropies (Smoot et al. 1992). Analyses of the COBE/DMR first year and first plus second years data have been performed using different techniques. Techniques based on the correlation function (Smoot et al. 1992, Scaramella and Vittorio 1993, Seljak and Bertschinger 1993, Bennett et al. 1994), on orthogonal functions on a sphere with the galactic plane removed (Wright et al. 1994b, Gorski et al. 1994), on RMS estimations (Smoot et al. 1994, Banday et al. 1994) and on the topology of the maps (Smoot et al. 1994, Torres 1994a,b, Torres et al. 1995) revealed the presence of signal in the analysed data and showed the degeneracy between the amplitude $Q_{rms-PS}$ and the index $n$ of the power spectrum of matter density fluctuations (higher values of $n$ require lower values of $Q_{rms-PS}$).

Since the announcement of the detection of anisotropies in the CMB temperature field by the COBE/DMR experiment, other experiments at different angular scales have also statistically detected CMB anisotropies in their data (see references in White, Scott and Silk 1994). Comparison of different experiments is required to confirm these detections. Ganga et al. 1993 found a strong statistical correlation between the 170 GHz partial sky survey by the FIRS experiment and the COBE/DMR first year data, indicating that the fluctuations observed by both experiments should correspond to the same anisotropies of the CMB. The detection of individual features opens new possibilities for comparison of the observations of different experiments. The first experiment for which the detection of individual, primordial features has been claimed is the Tenerife experiment (Hancock et al. 1994). Possible contamination of the anisotropies observed by the COBE/DMR experiment has been studied regarding systematic effects (Kogut et al. 1992), Galactic and extragalactic foregrounds (Bennett et al. 1992, 1993) and magnitude of the noise correlation (Lineweaver et al. 1994). Although none of these effects are significant, the detection of individual structures requires special care in relation to the level of noise present in the maps at different frequencies.

The idea of this work is first, to confirm the presence of cosmological signal in the first plus second year COBE/DMR data by using characteristics of the features that appear in the maps, and second, to determine the significance of the hot and cold spots as compared with noise. A description of the analysed data can be found in Section 2. Statistical analysis of the 53 GHz map is presented in Section 3; the method is explained in Subsection 3.1 and the results are given in Subsection 3.2. In Subsection 4.1 we describe the method used to estimate the significance of the observed spots as compared with noise and the results are given in Subsection 4.2. A summary of the main conclusions is given in Section 5.

## 2. Description of the Data

The COBE/DMR produced CMB temperature sky maps at three frequencies 31, 53 and 90 GHz. Observations are made by two different channels $A$ and $B$ at each frequency. Data of the two channels were combined in a sum $(A + B)/2$ and a difference $(A - B)/2$ map. The former is a combination of CMB signal and instrumental noise while the later contains only noise.

We analyse the first and second year data taken by the COBE/DMR experiment during 1990 and 1991. The analyzed maps ($(A + B)/2$ maps at different frequencies) have monopole and dipole (and optionally quadrupole) subtracted and a galactic cut of $|b| > 20°$ is carried out leaving 4016 pixels. The galactic cut does not introduce bias in our analysis as we consider independently spots appearing above and below the galactic cut. Moreover the realizations performed in this work are treated the same as the data. The radiometer noise lies in the range $[-1.39, 1.34]$ mK at 31 GHz, $[-0.36, 0.34]$ mK at 53 GHz and $[-0.52, 0.56]$ mK at 90 GHz with RMS 0.28 mK, 0.09 mK and 0.14 mK respectively. In order to improve the signal-to-noise ratio a Gaussian smoothing of $FWHM = 7°$ is performed, resulting in values of the signal-to-noise ratio of $1.1, 1.4$ and $1.13$ for the $31, 53$ and $90$ GHz data respectively. We searched for connected pixels (spots) above and below several thresholds $\nu$. The threshold is defined so that pixels above/below this level would have a temperature greater/lower or equal to $\nu \times RMS_{(A-B)/2}$. Figures 1a and 1b show the spots found at the 3 frequencies above/below levels 1/-1 and 2/-2 respectively. At level 1 (Figure 1a), six cold and eleven hot spots can be simultaneously observed at the three frequencies.

## 3. Statistical Analysis

### 3.1. Method

The purpose of this section is to set constraints on the amplitude of the power spectrum of the matter density fluctuations assuming a Harrison-Zeldovich power spectrum. We analyze the first plus second year data observed by the COBE/DMR experiment at 53 GHz as this is the channel with the best signal-to-noise ratio.

This analysis is based on Monte Carlo simulations made following the configuration of the DMR experiment. We simulate the CMB sky as seen by the DMR (beam filter given in Wright et al. 1994a) and create two maps, one including the noise of channel $A$ and the other one including the noise of channel $B$. After that, the sum and difference maps are created. Mean and dipole subtraction, galactic cut and smoothing are performed as indicated in the previous section (quadrupole subtraction is also performed to compare with results including the quadrupole).

We use a generalized $\chi^2$ statistic to determine the models favoured by the COBE/DMR data, considering the number of spots that appear at different thresholds and the fraction of the total area covered by them, $FA$. It should be noted that we are calculating these quantities as a function of threshold characterized by the dispersion of the noise map, as indicated in the previous section. This definition introduces a dependence of the $FA$ covered by the spots on the amplitude of the power spectrum.

The generalized $\chi^2$ method has been previously used by Torres et al. 1995 to set constraints on the amplitude of the power spectrum and the spectral index by analysing the genus in the first year COBE/DMR maps. The generalized $\chi^2$ for each realization $k$ is given by:

$$\chi_k^2 = \sum_{i=1}^{n_l} \sum_{j=1}^{n_l} (N_s^k(\nu_i) - N_s^d(\nu_i)) M_{ij}^{-1} (N_s^k(\nu_j) - N_s^d(\nu_j)) \,, \tag{1}$$

where $n_l$ is the number of thresholds ($\nu$) considered, $N_s^k(\nu_i)$ refers to the number of spots that appear at level $\nu_i$ in the $k$ realization and $N_s^d(\nu_i)$ is the number of spots found in the data at $\nu_i$. $M_{ij}$ is the covariance matrix defined by:

$$M_{ij} = \frac{1}{n_r} \sum_{k=1}^{n_r} (N_s^k(\nu_i) - <N_s(\nu_i)>)(N_s^k(\nu_j) - <N_s(\nu_j)>) \,, \tag{2}$$

where $n_r$ is the number of realizations, $n_r = 1000$ and the brackets represent the mean value averaging over the realizations. The $\chi^2$ for the fraction of the area occupied by the spots is defined in the same way by substituting $FA$ for $N_s$. We test models with values of the amplitude $Q_{rms-PS} = 0, 6, 9, 12, 15, 18, 21, 24, 27, 30$. In Figure 2, we present an example of the distribution of $\chi^2$ for three different models, calculated for the number of spots. As one can see from this figure, there is a significant statistical difference between different values of $Q_{rms-PS}$. For each of those models we first determine the mean value of the $\chi^2$ obtained by comparing with the COBE/DMR data, averaged over 1000 realizations, and select the model with the minimum value. The error bars associated with that model are calculated by assuming that each of the realizations of that model are now the data and repeating the process.

### 3.2. Results

The number of spots that a model produces in the map at different thresholds compared with what is observed in the first plus second year 53 GHz map gave a constraint on the amplitude of the power spectrum of $Q_{rms-PS} = 15^{+3}_{-6}\mu K$ at the 95% confidence level, assuming a Harrison-Zeldovich spectrum. By applying the same technique to the FA we got $Q_{rms-PS} = 18^{+3}_{-4}\mu K$, at the same confidence level. The same analysis is performed excluding the quadrupole, giving $18^{+5}_{-7}\mu K$ for the number of spots and $21 \pm 6\mu K$ for the fractional area, at the 95% confidence level. In agreement with previous analyses of the COBE/DMR two years data, the $Q_{rms-PS}$ values are larger when the quadrupole is subtracted, as a result of the low value of the observed quadrupole. The intervals overlap and more importantly the result confirms the presence of signal in the analyzed map of two years of observations. Therefore, our purpose in the next section is to estimate which of the spots observed in the maps have the lowest probability of being produced by noise and thus are likely real signal.

## 4. Significance of the Spots

### 4.1. Method

The aim of this section is to determine the significance of the spots present in the COBE/DMR maps in comparison with noise. For this purpose we study the distribution of several descriptors of spots in noise realizations and compare them with the corresponding values for the spots found in the data maps. The presence of identifiable spots in the data observed at the three frequencies is required. Once the spots found simultaneously in the three maps are selected, we concentrate in the 53 GHz map as it is the one with the highest signal-to-noise ratio.

The number of spots found in coincidence at the three frequencies is a function of the required number of common pixels per spot. We performed 1000 noise realizations following the prescription indicated in Subsection 3.1. Each realization of simulated noise maps at the three frequencies are compared and the number of triple coincidences are counted as a function of the required number of common pixels/spot. The coincidence-finding procedure begins by comparing the 53 and 31 GHz maps. Next the selected common spots at 53 GHz are compared with the spots at 90 GHz. The final result is a number of cold and hot spots appearing at 53 GHz that have been found in coincidence at the other two frequencies and for which we determine the area and the mean signal-to-noise ratio as will be indicated below. Table 1 presents the percentage of noise realizations as a function of the number of common spots found below/above threshold -1/1 imposing different number of common pixels per spot.

Tables 2a and 2b present the common cold and hot spots found in the COBE/DMR two year data below/above threshold -1/1. Spots marked with an asterisk were found imposing the condition of having at least five pixels in common per spot and the ones marked with two asterisks were found in common at the three frequencies with seven or more common pixels per spot. Although we are removing 20 degrees above and below the galactic plane, it is still very possible that part of the hot spots can be produced by galactic contamination. The second and third spots in Table 2a are situated in the Ophiuchus complex. The large spot situated at $(l, b) = (176°, -42°)$ is close to the Orion complex and therefore part of the signal may be produced by the Galaxy. Not considering these spots, we are left with eight hot spots and six cold spots when no restrictions are imposed regarding the number of common pixels per spot. Considering only common spots with more than 4 and 6 common pixels we have 3 hot, 4 cold spots and 3 hot, 3 cold spots respectively. Comparing these numbers with the results of noise realizations presented in Table 1 one can see that a noise realization is very unlikely to produce the observed distribution. The percentage of noise realizations that would produce the same number of observed common hot and cold spots is less than 12.5% not imposing restrictions in the number of common pixels per spot and less than 2.5% and 0.7% considering only spots with more than 4 and 6 pixels in common respectively. Therefore, the data that we are analysing presents some spots for which a non-noise origin could be attributed.

Once the spots are selected above different thresholds in the data maps as well as in the simulated maps (resulting from realizations performed in the way indicated in the previous section) we determine the position of the barycenter (galactic coordinates) and calculate the area $A$ and the signal-to-noise ratio $S/N$ (as we will see, the $S/N$ defined for the spots should not be confused with the signal-to-noise defined for the map in the previous section). The area is defined by the number of connected pixels, where each pixel has a size of $2°.6 \times 2°.6$. The signal-to-noise ratio of a spot is given by the average of the signal-to-noise ratio of the connected pixels inside the spot $n_c$, that is, by the average of the ratio of the temperature $T_i$ over the ratio of the channel dependent instrument noise dispersion per observation $\sigma_s$ over the square root of the number of observations taken in each pixel $N_i$,

$$S/N \equiv \frac{1}{n_c} \left| \sum_{i=1}^{n_c} \frac{T_i}{\frac{\sigma_s}{\sqrt{N_i}}} \right| . \qquad (3)$$

As indicated above we performed 1000 noise realizations and imposed different criteria to select the common spots observed simultaneously at the three frequencies. We calculate the distribution of the area and the mean signal-to-noise. Seventeen thresholds, in between -4 and 4, are analysed. Comparison of the values found for each individual spot seen simultaneously in the three data maps at a fixed threshold with the noise error bars will give their significance level and is presented in the following subsection.

### 4.2. Results

Tables 2.a and 2.b present the characteristics of the common spots as they appear in the 53 GHz map below/above threshold -1/1. The position of the barycenter is given in galactic coordinates (degrees) in the first two columns. The third and fifth columns give the area and the $S/N$ of the spots. The probability of having spots produced by noise with area ($A$) and signal-to-noise ($S/N$) lower than the observed value is given in the fourth (sixth) column. As an example, the probability distribution function of the area and of the $S/N$ can be seen in Figures 3.a and 3.b respectively, in the case that no restrictions are imposed on the required number of common pixels per spot. The highest level at which the spot is observed is given in the seventh column of Tables 2.a and 2.b. The mean value of the area that the spots occupy at the highest/lowest observed threshold is of approximately eight pixels ($\sim 7°$ in extension), that is $\sim 700 h^{-1}$ Mpc.

As indicated above, a realization of noise with the characteristics of the COBE/DMR noise at 33, 53 and 90 GHz for which we find 3 cold and 3 hot spots at 53 GHz in coincidence with more than 6 pixels in common per spot is very unlikely. From Table 2a one can see that among these cold spots the spot situated at $(l, b) = (-21°, -45°)$ has an area and a $S/N$ with probability less than 1% of being produced by noise. The cold spot at $(l, b) = (-99°, 57°)$ has $A$ and $S/N$ with probability less than 5% of being produced by noise. In Table 2b we find just one hot spot at $(l, b) = (-81°, -33°)$ with area and signal-to-noise values with probability higher than 95% of being produced by cosmological signal (we are not considering the hot spots that could be contaminated by Galactic signal). The results do not change when one considers spots found with more than 4 pixels in common per spot. Not imposing restrictions in the number of common pixels per spot there is one more cold spot that has area and $S/N$ with probability higher than 95% of being produced by cosmological signal $(l, b) = (85°, 40°)$. This spot does not have at least 5 pixels in common in the three maps. The selected hot spot at $(l, b) = (-81°, -33°)$ has in this case probability less than 1% of being caused by noise.

We have also checked how different smoothing angles affect the final results. As the smoothing angle grows the observed data have higher probability of being a noise realization. Requiring more than four common pixels per spot, the probability that the observed realization corresponds to noise grows from less than 2.5% to $\sim 20\%$ for $FWHM = 20°$ for the distribution of number of hot and cold spots. No new common spots were found when analysing the data at different smoothing angles. The selected spots appear with probability higher than 95% of not being produced by noise when looking at the area and the signal-to-noise. In addition to these spots, the cold spots at $(l, b) = (73°, -29°), (81.5°, -59°)$ appear as only one when $FWHM \geq 15°$ and with probability less than 5% of being caused by noise. The cold spot at $(l, b) = (85°, 40°)$ has high probability of not being produced by noise when larger smoothing angles are considered.

## 5. Conclusions

The presence of cosmological signal in the data observed during the first two years by the COBE/DMR experiment have been analyzed considering the characteristics of the features that appear in the maps. We performed a statistical analysis based on Monte Carlo simulations of a Harrison Zeldovich power spectrum to set constraints on the amplitude of the power spectrum. A $\chi^2$ statistic was calculated to compare the number of spots and the fraction of the area occupied by them assuming different models, with the first plus second year COBE/DMR data. From the number of spots we obtained, at the 95% confidence level, $Q_{rms-PS} = 15^{+3}_{-6}, 18^{+5}_{-7} \mu K$ and from the fractional area $Q_{rms-PS} = 18^{+3}_{-4}, 21 \pm 6 \mu K$, including and not including the quadrupole respectively. There have been previous analyses of the first year of data of the COBE/DMR experiment based on the characteristics of the features observed at different thresholds (Smoot et al. 1994, Torres 1994a,b, Torres et al. 1995). They obtained values of $Q_{rms-PS}$, with the spectral index forced to unity, which fall within the error bars of our analysis. The $RMS$ of the first plus second year COBE/DMR data has been calculated by Banday et al. 1994. The $1\sigma$ intervals of the $Q_{rms-PS}$, subtracting and not subtracting the quadrupole, overlap with our results.

We have also determined the significance level of the hot and cold spots present in the first plus second year COBE/DMR maps as compared with noise. We used a statistical method based on Monte Carlo realizations of pure noise. The descriptors of the spots used for that analysis are the area and the signal-to-noise ratio. The presence of the spots in the three maps (at the three different frequencies) was required. The condition of different number of common pixels per spot required has been studied.

The second and seventh hot spots that appear in Table 2b have an area and $S/N$ with probability higher than 95% of not being noise. The former spot is situated above the Galactic center on the Ophiuchus loop which is a known feature of Galactic structure. The second spot is covering part of the Orion complex. Therefore these spots are very likely contaminated by Galactic emission that could still be affecting the analysed maps. The fact that their intensity decreases with the frequency is another indication of their probably Galactic origin (although the spots have a higher intensity in the 53 GHz map than in the 31 GHz data; however, the level of noise at 31 GHz is larger than at 53 GHz). We did not include these spots in the analysis of the significance of the spots. The third hot spot appearing in Table 2b was also excluded from our analysis as it could also be part of the Ophiuchus loop appearing as a very small spot below $b = -20°$.

As the number of common pixels per spot increases the chances of finding spots produced by noise in coincidence at the three frequencies decreases. These results were presented in Table 1. The significance of the observed spots estimated from the area and the signal-to-noise compared with the distributions of these two quantities are given in Tables 2a and 2b. The probabilities of the spots of being produced by signal do not change significantly with the required number of common pixels per spot. There is one cold spot that clearly has a probability of 99% of being produced by cosmological signal. That spot is located at $(l, b) \approx (-21°, -45°)$. Considering spots with probability less than 5% of being caused by noise we find a cold spot at $(l, b) \approx (-99°, 57°)$ and a hot spot located at $(l, b) \approx (-81°, -33°)$. The probability of generating a realization of noise spots as the one observed increases with the smoothing angle. As the results are not quantitatively modified, to be conservative we would point out at the 95% confidence level the three spots (two cold and one hot) indicated above.

An analysis of the spots observed in the first year COBE/DMR data is presented in Torres 1994b. The level at which the spots are observed, as well as the area and the eccentricity are calculated. All those spots are also present in our data. In Torres 1994b there is no estimation of the noise contribution to the individual spots and the main conclusion is the compatibility of the data with a Harrison-Zeldovich spectrum with amplitude $Q = 16 \, \mu K$. The Wiener filter technique has been applied by Bunn et al. 1994 to the Reduced Galaxy linear combination of the first year COBE/DMR sky maps. This technique requires the assumption of a certain power spectrum of the matter density fluctuations. Assuming a spectral index $n = 1.0$ and the corresponding spectral amplitude (Seljack and Bertschinger 1993) they performed constrained realizations of the CMB anisotropy. They observe two hot spots $(l, b) = (-85°, -36°), (55°, 65°)$ and a large cold spot at $(l, b) = (-100°, 50°)$ that appear highly significant in the simulated maps. All those spots are considered in our analysis as they appear in coincidence in the 31, 53 and 90 GHz two years maps. As indicated above, the hot spot located at $(l, b) = (55°, 65°)$ is very likely produced by the Galaxy (the Reduced Galaxy map used by Bunn et al. 1994 may still have galactic contamination). The other hot spot is one of the spots pointed out in this work with a probability lower than 5% of being produced by noise. The large cold spot observed at $(l, b) = (-100°, 50°)$ corresponds to the first one in Table 2a. In our analysis this spot also have a high probability of being caused by cosmological signal (higher than 95%). There is a remarkable agreement between the results found by these two works.

Up to now, comparison between results from different experiments has been performed by statistical analysis of the observed fluctuations. Confirmation of those results can now be done by observation of the individual spots detected by the different experiments. Moreover, all CMB anisotropy experiments are constrained to observe the same sky that is, the same realization of the stochastic temperature fluctuation field. Information on the presence of a hot/cold spot detected by an experiment in the region sampled by another CMB anisotropy experiment can change the interpretation of the cosmological implications inferred by comparing both experiments. New techniques of data analysis based on that principle are starting to appear in the literature (Atrio-Barandela, Cayón and Silk 1995). The analysis of the four years of COBE/DMR data will improve the significance of the results.

We acknowledge the excellent work of those contributing to the COBE/DMR. We would also like to thank A. Banday, A. Kogut, C. Lineweaver, E. Martínez-González, and L. Tenorio for helpful discussions. This work was supported in part at LBL through DOE Contract DOE-AC-03-76SF0098. L.C. acknowledges the support of a MEC/Fulbright fellowship (FU93 13924591). COBE is supported by the Office of Space Sciences of NASA Headquarters.

| $N_{common}$ | $n_{req} = 7$ | $n_{req} = 5$ | $n_{req} = 1$ |
|:---:|:---:|:---:|:---:|
| 1 | 20.0/24.0 | 36.0/36.0 | 04.0/04.0 |
| 2 | 03.0/03.0 | 14.0/12.0 | 11.0/11.0 |
| 3 | 00.1/00.7 | 02.4/01.6 | 21.0/19.0 |
| 4 | 00.0/00.0 | 00.5/00.2 | 20.5/23.0 |
| 5 | 00.0/00.0 | 00.0/00.0 | 19.0/18.0 |
| 6 | 00.0/00.0 | 00.0/00.0 | 12.5/13.0 |
| more than 7 | 00.0/00.0 | 00.0/00.0 | 12.0/11.0 |

Table 1: Percentage of noise realizations as a function of the number of common spots found at the three frequencies below/above threshold -1/1. Column one gives the number of common spots $N_{common}$. Columns 2,3 and 4 present the percentage of realizations imposing different number of common pixels/spot $n_{req} = 1, 5, 7$. The first number in each of the last three columns gives the percentage of realizations as a function of the common cold spots and the second number gives the same quantity as a function of the common hot spots.

| $l(°)$ | $b(°)$ | $A(\#pixels)$ | $P(A)\%$ | $S/N$ | $P(S/N)\%$ | $\nu_{max}$ |
|---|---|---|---|---|---|---|
| ** -99 | 57 | 259 | 100/100/100 | 1.97 | 97/97/98 | -4.0 |
| 85 | 40 | 82 | 97 | 2.23 | 100 | -2.5 |
| ** -21 | -45 | 162 | 100/100/100 | 2.50 | 100/100/100 | -3.5 |
| ** 73 | -29 | 35 | 77/64/51 | 2.30 | 100/100/100 | -3.5 |
| * 81.5 | -59 | 127 | 100/99 | 1.66 | 80/78 | -2.0 |
| -86 | 33 | 9 | 22 | 1.44 | 45 | -1.0 |

Table 2a: Spots found below threshold −1 in the 53 GHz map, that appear simultaneously in the three DMR maps, at 31, 53 and 90 GHz. The first two columns give the position of the barycenter in galactic coordinates, $l,b$ in degrees. The third column shows the area, $A$, of the spots and the fourth column indicates the probability of finding spots created by pure noise with area smaller than the area of the spot found in the data, $p(A)$. The signal-to-noise of the spots, $S/N$ is given in column five and the probability of having spots created by noise with $S/N$ smaller than the signal-to-noise of the data is given in column six. Spots marked with an asterisk have been found in coincidence with more than 4 pixels in common per spot. Two asterisks indicate the common spots with more than 6 pixels in common at the three frequencies. In these cases the second number given in columns fourth and sixth indicates the probability obtained form noise realizations imposing a minimum of 5 common pixels per spot and the third number gives the same quantity obtained from noise realizations imposing a minimum of 7 common pixels per spot. Column seven shows the highest level at which the spots are observed.

| $l(°)$ | $b(°)$ | $A(\#pixels)$ | $P(A)\%$ | $S/N$ | $P(S/N)\%$ | $\nu_{max}$ |
|---|---|---|---|---|---|---|
| -24 | 51 | 29 | 69 | 1.86 | 95 | 2.5 |
| ** 41 | 71 | 375 | 100/100/100 | 1.99 | 98/99/98 | 4.0 |
| 12.5 | -24 | 10 | 24 | 1.88 | 96 | 2.0 |
| ** 46 | -32 | 53 | 90/83/74 | 1.65 | 81/78/77 | 2.5 |
| 96 | -28 | 27 | 66 | 1.88 | 96 | 2.5 |
| 120 | -36.6 | 56 | 91 | 1.70 | 86 | 2.5 |
| ** 176 | -42 | 100 | 99/98/96 | 1.91 | 97/97/96 | 3.0 |
| ** 172 | 29 | 33 | 73/57/46 | 2.25 | 100/100/100 | 3.0 |
| ** -81 | -33 | 105 | 99/98/97 | 2.40 | 100/100/100 | 3.0 |
| 141 | -73.5 | 69 | 95 | 1.56 | 69 | 2.0 |
| -82 | -58 | 8 | 16 | 2.08 | 99 | 1.0 |

Table 2b: Spots found above threshold 1 in the 53 GHz map, that appear simultaneously in the three DMR maps, at 31, 53 and 90 GHz. First two columns give the position of the barycenter in galactic coordinates, l,b in degrees. The third column shows the area, $A$, of the spots and the fourth column indicates the probability of finding spots created by pure noise with area smaller than the area of the spot found in the data, $p(A)$. The signal-to-noise of the spots, $S/N$ is given in column five and the probability of having spots created by noise with $S/N$ smaller than the signal-to-noise of the data is given in column six. Spots marked with an asterisk have been found in coincidence with more than 4 pixels in common per spot. Two asterisks indicate the common spots with more than 6 pixels in common at the three frequencies. In these cases the second number given in columns fourth and sixth indicates the probability obtained form noise realizations imposing a minimum of 5 common pixels per spot and the third number gives the same quantity obtained from noise realizations imposing a minimum of 7 common pixels per spot. Column seven shows the highest level at which the spots are observed.

Fig. 1a.— COBE/DMR 2 years maps at 31, 53 and 90 GHz showing the spots found at $|\nu| > 1$. The maps are presented in Polar or Sterographic projection. Positive galactic latitudes are on the left and negative are on the right. Galactic longitude grows from 0° to 360° in the clockwise direction in the north and in the opposite direction in the south.

Fig. 1b.— Polar representation of the COBE/DMR 2 years maps at 31, 53 and 90 GHz showing the spots found at $|\nu| > 2$.

Fig. 2.— $\chi^2$ distribution for a 1000 realizations of noise (solid line) and of Harrison-Zeldovich power spectrum fluctuations with $Q_{rms-PS}$ = 15 (dotted line), 27 (dashed line) $\mu K$, as compared with the first plus second year COBE/DMR data, by analysing the number of spots that appear at several thresholds.

Fig. 3a.— Distribution of the area of the spots observed above/below level 1/-1 (solid/ dashed line) in 1000 noise realizations. No restrictions in the number of common pixels per spot.

Fig. 3b.— Distribution of the signal-to-noise of the spots observed above/below level 1/-1 (solid/ dashed line) in 1000 noise realizations. No restrictions in the number of common pixels per spot.